\documentclass[12pt]{iopart}

\RequirePackage{lineno}
\usepackage{graphicx}% Include figure files 
\usepackage{dcolumn}% Align table columns on decimal point 
\usepackage{epstopdf}
\usepackage{color}
\usepackage{amssymb}
\usepackage{bm}% bold math

\def\bea {\begin{eqnarray}}
\def\eea {\end{eqnarray}}

\def\be {\begin{equation}}
\def\ee {\end{equation}}

\newcommand{\Npart}{$\langle N_{\rm part} \rangle$~}

\newcommand{\sNN}{$\sqrt{s_{\rm {NN}}}$}    

\newcommand{\wch}{$\omega_{\rm {ch}}$}
\newcommand{\acceptance}{$|\eta|<0.5$ and $0.2 < p_{\rm T}< 2.0$~GeV/c}
\begin{document}

\title{Fluctuations in Charged Particle Multiplicities in Relativistic
  Heavy-Ion Collisions}

\author{Maitreyee Mukherjee, Sumit Basu, Subikash Choudhury \& Tapan K. Nayak}

\address{Variable Energy Cyclotron Centre, Kolkata - 700064, India}
\ead{maitreyee.mukherjee@cern.ch}
\vspace{10pt}

\date{\today}
%\linenumbers
\begin{abstract}

Multiplicity distributions of charged particles and their
event-by-event fluctuations have been compiled for relativistic
heavy-ion collisions 
from the available experimental data at Brookhaven National Laboratory
and CERN and also by the use of an event generator.
Multiplicity
fluctuations are sensitive to QCD phase transition and to the presence
of critical point in the QCD phase diagram. In addition, multiplicity
fluctuations provide baselines for other event-by-event measurements. 
Multiplicity fluctuation expressed in terms of the scaled variance of the
multiplicity distribution is an
intensive quantity, but is sensitive to the volume fluctuation of the system.
The importance of the
choice of narrow centrality bins and the corrections of centrality
bin width effect for controlling volume fluctuations have been discussed. 
It is observed that the mean and width of the multiplicity distributions monotonically
increase as a function of increasing centrality at all collision
energies, whereas the
multiplicity fluctuations show minimal
variations with centrality. The beam energy dependence shows that the
multiplicity fluctuations have a slow rise at lower collision
energies and remain constant at higher energies.

\end{abstract}
%\maketitle

\section{Introduction}

One of the basic advantages of the heavy-ion collisions at 
relativistic energies is the production of large number of particles in each
event, which facilitates the event-by-event study of several
observables. The major physics goals at these high energies are to understand the
nature of phase transition from normal hadronic 
matter to a phase of quark-gluon plasma (QGP). This topic has been of
tremendous interest over last four decades, both in terms of
theoretical studies and large scale experiments. 
Dedicated experiments have been performed at
the Alternating Gradient Synchrotron (AGS) and 
Relativistic Heavy Ion Collider (RHIC) at 
Brookhaven National Laboratory, and the Super Proton Synchrotron (SPS) and
the Large Hadron Collider (LHC)
at CERN to search as well as study the QGP phase. 
These experiments explore the QCD phase transition.
The fluctuations of experimentally accessible quantities, such as 
particle multiplicity, mean transverse momentum, temperature,
particle ratios, and other global observables are related to the 
thermodynamic properties of the system, such as the entropy, specific heat,
chemical potential and matter 
compressibility~\cite{shuryak,rajagopal,stephanov,marek}. 
Fluctuations of these quantities on an event-by-event basis
have been used as basic tools for understanding the particle production
mechanisms, the nature of the phase transition and critical
fluctuations at the QCD phase boundary.
A non-monotonic behaviour of the fluctuations as a function of
collision centrality and energy may 
signal the onset of deconfinement, and can be effectively
used to probe the critical point in the QCD phase diagram. 

Theoretical models, based on lattice QCD, reveal that
at vanishing baryon chemical potential ($\mu_{B}$), the
transition from QGP to hadron gas is a  
crossover, whereas at large $\mu_{B}$, the phase
transition is of first order~\cite{aoki}.
Experimental observables at SPS and RHIC energies may point to the onset
of deconfinement and a hint of 
first order phase transition has been indicated~\cite{marek,marek1,marek2,indication,star_qcd}.
First order phase transitions can lead to large 
density fluctuations resulting in bubble or droplet formation and hot
spots~\cite{heiselberg,bubble1,bubble2,bialas,baym}, which give rise
to large multiplicity fluctuations in a given rapidity
interval. The local multiplicity fluctuations have been predicted as a
signature of critical hadronisation at RHIC and LHC energies~\cite{hwa}.
Measurements at the vanishing $\mu_{\rm B}$
at  LHC energies will be important as
one can accurately calculate several quantities and their
fluctuations. 

The multiplicity of produced particles is an important quantity which 
characterises the system produced in heavy-ion 
collisions. Consequently, multiplicity and its fluctuation has an effect on all
other measurements. Multiplicity fluctuations have been characterised by the scaled variances of 
the multiplicity distributions, defined as,
\begin{eqnarray}
\omega=\frac{\sigma^2}{\mu},
\end{eqnarray}
where $\mu$ and $\sigma^2$ are the mean and 
variance of the multiplicity distribution, respectively.
Multiplicity fluctuations have been reported by E802 experiment~\cite{e802} at AGS,
WA98~\cite{wa98}, NA49~\cite{na49,na49-Ryb}, NA61~\cite{na61a,na61b} and CERES~\cite{sako}
experiments at SPS, and
the PHENIX experiment~\cite{Phenix} at RHIC. The nature of 
the multiplicity distributions as a function of centrality and beam 
energy has been compared to statistical and available model 
calculations. These results have generated a great deal of  
interest~\cite{Phenix, Begun, Jeon, Zozulya, Becattini}.

Multiplicity fluctuations have contributions from
statistical (random) components as well as those which have dynamical (deterministic) origin. 
The statistical components have contributions from the
choice of centrality, fluctuation in impact parameter or
number of participants, finite particle multiplicity, effect of
limited acceptance of the detectors, fluctuations in the number of
primary collisions, effect of rescattering,
etc.~\cite{rajagopal,baym,hwa2}. The statistical components of the multiplicity 
fluctuations have direct impact on the fluctuations in other measured quantities. 
The dynamical part of the fluctuations contain interesting physics
associated with the collision, which include time evolution of
fluctuations at different stages of the collision, hydrodynamic
expansion, hadronisation and freeze-out.
In order to extract the dynamical part of the fluctuations, the
contribution to multiplicity from statistical components has to be well
understood. We discuss the methods for controlling geometrical
fluctuations so that dynamical fluctuations, if present, become more prominent.

In this article, we present a study of charged particle multiplicity fluctuations as a function
of centrality and beam-energy for Au+Au collisions for the Beam Energy
Scan (BES) energies at RHIC (from \sNN~=~7.7~GeV to 200 GeV) and Pb+Pb
collisions at LHC-energy (\sNN~=~2.76~TeV) from the
available experimental data as well as using different modes of the
AMPT (A Multi-Phase Transport)
model~\cite{ampt}.
In the next section, we present different model
settings of AMPT. In
section III, we discuss the method of centrality selection for
fluctuation studies and the centrality bin width corrections.
Multiplicity distributions for all the collision energies are presented
in section IV.  The results of multiplicity fluctuations are
given in section V. A discussion of the results is presented in
section VI and the paper is summarised in section VII. 

\section{AMPT Model}

The AMPT model~\cite{ampt,ampt2,ampt3} is used as a guidance for
obtaining multiplicity distributions and fluctuations wherever the
experimental data are not available. The model
consists of four
main components: the initial conditions, partonic
interactions, the conversion from the partonic to the hadronic matter, and
hadronic interactions. 
The  model provides two modes: Default and String Melting (SM)~\cite{ampt}. 
In both the cases, the  initial conditions are taken from HIJING~\cite{hijing} with 
two Wood-Saxon type radial density profile of the colliding nuclei.
The multiple scattering among the nucleons of two heavy ion nuclei are
governed by the eikonal formalism. In the default mode, energetic
partons recombine and hadrons are produced via string fragmentation. The string fragmentation takes place 
via the Lund string fragmentation function, given by,
\begin{eqnarray}
f(z) \propto~ z^{-1}(1-z)^{a}exp(-\frac{b m^2_{T}}{z}).
\end{eqnarray}
Interactions of the produced hadrons
are described by A Relativistic Transport model (ART).

In the SM mode, the strings produced from HIJING are
decomposed into partons which are fed into the parton cascade along
with the minijet partons. The partonic matter coalesce to produce
hadrons, and the hadronic interactions are
subsequently modelled using ART. 
While the Default mode describes the collision evolution in
terms of strings and minijets followed by string fragmentation, the
SM mode includes a fully partonic QGP phase that
hadronises through quark coalescence. 

For both the modes, Boltzmann equations are 
solved using Zhang's parton cascade (ZPC) 
with total parton elastic scattering cross section, 
\begin{eqnarray}
\sigma_{gg} = \frac{9\pi \alpha^2_s}{2\mu^2} \frac{1}{1+\mu^2/s} \approx~\frac{9\pi \alpha^2_s}{2\mu^2},
\end{eqnarray}
where $\alpha_s$ is the strong coupling constant, $s$ and $t$ are the
Mandelstam variables and $\mu$ 
is the Debye screening mass. Here $a$, $b$ (fragmentation parameters),
 $\alpha_s$ and $\mu$ are the
key deciding factors for multiplicity yield at particular beam
energy. The values are taken as 2.2, 0.5, 0.47 
and 1.8~fm$^{-1}$ respectively, corresponding to total parton
elastic cross section  $\sigma_{gg}$=10mb.
The mean values of multiplicities are found to match to the
experimental data with these tunings. The AMPT model,
therefore, provides a convenient way to investigate a
variety of observables with the default and SM modes.

\section{Centrality selection and centrality bin width correction}

The particle production mechanisms are expected to be dependent on the
collision energy as well as the centrality of the collision. For most
of the analysis, it is important to consider proper centrality window
so that fluctuations because of the selection are minimised.
Centrality of a collision is characterised by the impact parameter ($b$) of
the collision or equivalently the average number of participating nucleons (\Npart). 
In an experimental scenario it is not possible to access these two
quantities, so charged particle multiplicities within a given rapidity
range or energy depositions by calorimeters are used. 
In a model dependent way, the connections of these experimental
quantities to $b$ or \Npart~are made.
This is indeed needed in order to connect any measured quantity 
with theoretical calculations and to compare them with measurements from other experiments.

\begin{figure}[tbp]
\begin{center}
\includegraphics[width=0.52\textwidth]{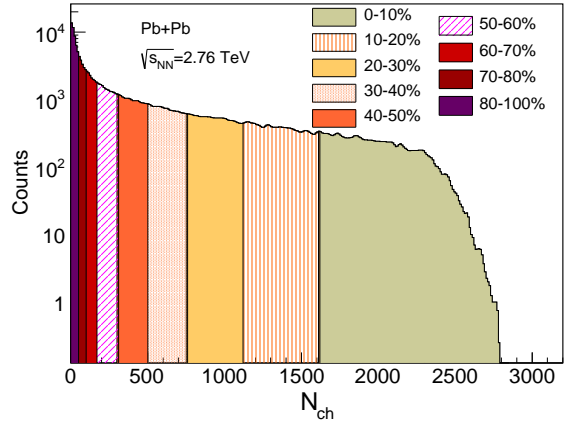}
\caption{(colour online) An example of centrality selection from minimum-bias
distribution of charged particles generated with SM mode of AMPT for
  Pb+Pb collisions at  \sNN~=~2.76~TeV for 
$2.0 < |\eta| <
  3.0$ and $0.2 < p_{\rm T}< 2.0$~GeV/c.
}
\label{fig1}
\end{center}
\end{figure}

In the present study, centrality is selected using 
the minimum bias distribution of charged particles in the forward
pseudorapidity~($\eta$) range of  $2.0 < |\eta| < 3.0$, and the
multiplicity fluctuations are calculated in the central $\eta$-range
($|\eta| < 0.5$).  Thus the two $\eta$-ranges are very distinct and
the fluctuation results are unbiased.  
As an example of centrality selection procedure, in Fig.~\ref{fig1} we present the 
minimum bias charged particle multiplicity distribution within $2.0 < |\eta| < 3.0$
and  transverse momentum ($p_{\rm T}$) range of $0.2 < p_{\rm T}<
2.0$~GeV/c in Pb+Pb collisions at \sNN~=~2.76~TeV obtained from AMPT model.
Depending on the centrality selection requirement, the area under the
curve is divided into centrality percentiles.
The shaded regions in the figure show 
selections in 10\% centrality cross-section bins (20\% bin is shown for most peripheral
collisions).
For experimental data, centralities are selected by 
Glauber model fits to the minimum-bias distributions of charged 
particles~\cite{ALICEcentrality}. 

Selection of narrow centrality window is essential for any fluctuation
study. For multiplicity fluctuations, this can be understood in terms 
of a simple participant model. The number of produced particles ($N$) 
in a collision depends on \Npart and the number of collisions 
suffered by each particle. Mathematically this can be expressed as 
\begin{eqnarray}
N = \sum\limits_{i=1}^{N_{\rm part}}  n_i 
\end{eqnarray}
where $n_i$ is the number of particles produced in the $\eta$-window
of the detector by the $i^{th}$-participant. On an average, the mean value 
of $n_i$ is the 
ratio of the average multiplicity in the detector coverage to the 
average number of participants, 
i.e., $\langle n \rangle  = \langle N \rangle / \langle N_{\rm part}
\rangle$. Thus the fluctuation in particle multiplicity is directly 
related to the fluctuation in \Npart.  In order to infer 
dynamical fluctuations arising from various 
physics processes one has to make sure that the fluctuations in \Npart~are minimal. 

\begin{figure}[tbp]
\begin{center}
\includegraphics[width=0.55\textwidth]{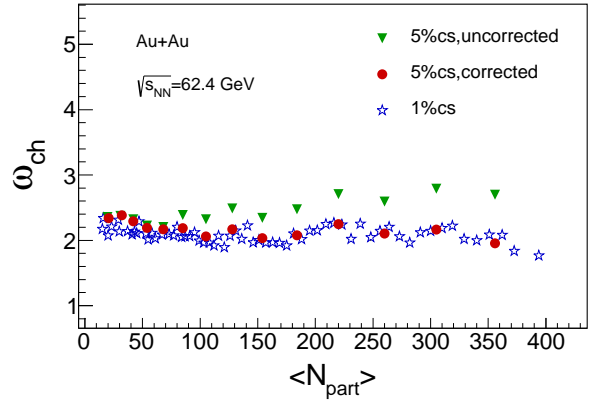}
\caption{(colour online) 
Effect of centrality bin width correction on
scaled variances ($\omega_{ch}$) is shown for 
choosing 5\% centrality bins. Results for
Au+Au collisions at $\sqrt{s_{NN}}=62.4$~GeV using the default mode of 
AMPT as a function of \Npart shows that after the correction 5\% centrality bins are similar
to those obtained for 1\% centrality bins. 
}
\label{fig2}
\end{center}
\end{figure}

Selection of narrow centrality bins helps to get rid of inherent 
fluctuations within a centrality bin. The inherent fluctuations are
intrinsic fluctuations arising from the difference in geometry even
within the centrality bin. A centrality bin scans a range of charged
particle multiplicity with different cross sections. This introduces
geometrical fluctuations which need to be controlled. 
Choosing very narrow centrality window minimises the geometrical fluctuations.
But it may not be always 
possible to present the results in such narrow bins, mainly because of lack of 
statistics and also because of centrality resolution of detectors used.
It is desirable to choose somewhat wider centrality bins,
such as 5\% or 10\% of the total cross sections. But these choices
introduce inherent fluctuations which need to be corrected.
This is done by taking the weighted average of the observables, such as,
\begin{equation}
X = \frac{\sum_{i}  n_{i}X_{i}}{\sum_{i}n_{i}},
\end{equation}
where the index $i$ runs over each multiplicity bin, 
$X_{i}$ represents various moments for the $i$-th bin, and 
$n_{i}$ is the number of events in the $i$-th multiplicity bin. 
$\sum_{i}n_{i} = N$ is the total number of events in the centrality
bin. 

This is demonstrated in Fig.~\ref{fig2} in terms of centrality
dependence of scaled variance of
the multiplicity distributions for Au+Au collisions at 
$\sqrt{s_{NN}}=62.4$~GeV with the generated events from the default 
version of AMPT. Three sets of \wch~values are presented. The values
of \wch~obtained with 5\% centrality bins are much larger compared to
the ones with 1\% centrality bins. This variation comes because of
wide 5\% bins. 
After making the correction of the bin width effect, the fluctuations for 
the 5\% cross section bins  reduce by close to $\sim 23\%$ and $\sim
8\%$, respectively 
for central and peripheral collisions, and almost coincide with that of the 
1\% cross section bin. No centrality bin width dependence is observed 
after employing the correction. Thus 
by choosing narrow bins in centrality and making centrality bin width
correction within each centrality window, the volume fluctuations are minimised. 

\section{Multiplicity distributions}

\begin{figure}[tbp]
\begin{center}
\includegraphics[width=0.52\textwidth]{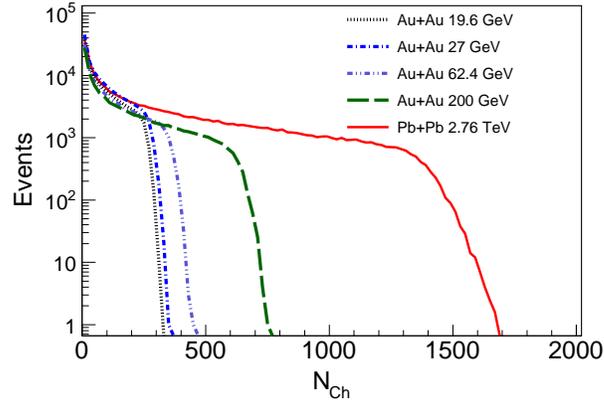}
\caption{(colour online) Minimum-bias distributions for charged particles 
for Au+Au collisions at 19.6, 27, 62.4 and 200 GeV using default
AMPT model and for Pb+Pb
collisions at \sNN~=~2.76~TeV, obtained
using SM version of AMPT within \acceptance.}
\label{multiplicity}
\end{center}
\end{figure}

Particle multiplicity distributions for different beam energies and
collision centralities help to understand the mechanisms of particle 
production and constrain various models.
Figure~\ref{multiplicity} shows minimum bias charged particle
multiplicity distributions for \acceptance~in Au+Au collisions at 
$\sqrt{s_{NN}}= $19.6 GeV, 27 GeV, 62.4~GeV and 200~GeV, 
using the default mode of AMPT, and Pb+Pb collisions at \sNN~=~2.76~TeV using
the SM mode of AMPT. As seen from the figure, each distribution gives
the maximum extent of the multiplicity for a given collision
energy for a given number of events. The maximum extent is larger for larger collision energy. 

\begin{figure}[tbp]
\begin{center}
\includegraphics[width=0.58\textwidth]{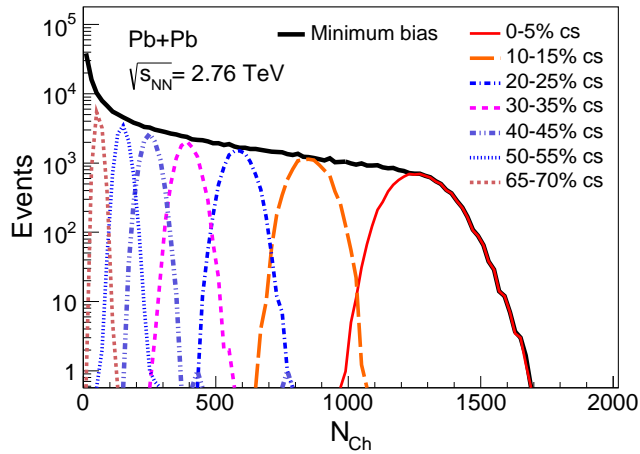}
\caption{(colour online) Charged particle multiplicity distributions
  for different centralities for Pb+Pb collisions at
\sNN~=~2.76~TeV using SM mode of AMPT model within \acceptance.}
\label{convolution}
\end{center}
\end{figure}

The minimum bias multiplicity distribution is a 
convolution of multiplicity distributions with different centrality bins. This is 
illustrated in Fig.~\ref{convolution} for charged 
particle multiplicity distributions in case of 
Pb+Pb collisions at \sNN~= 2.76 TeV from the SM mode of AMPT model. 
Minimum bias distribution as well as distributions at 
different centrality bins are presented.

\begin{figure}[tbp]
\begin{center}
\includegraphics[height=0.5\textheight,width=0.55\textwidth]{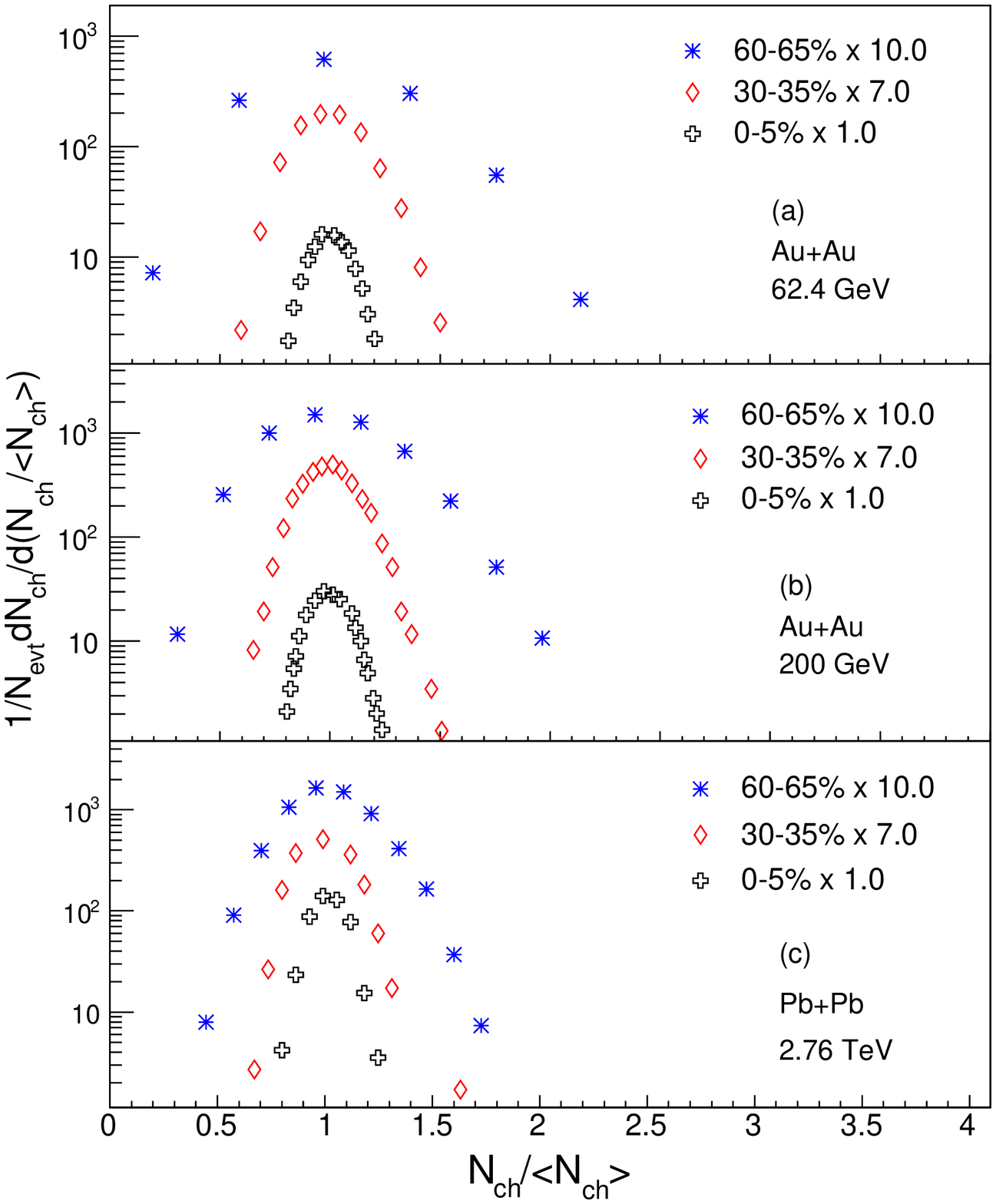}
\caption{(colour online) Scaled multiplicity distributions of charged
  particles for centralities corresponding to 0-5\%, 30-35\% and
  60-65\% cross sections within \acceptance~using 
  default AMPT for Au+Au collisions at (a) \sNN~=~62.4~GeV, (b)
  \sNN~=~200~GeV, and using SM mode of AMPT for  Pb+Pb collisions at
  (c) \sNN~=~2.76~TeV.
  The charged particle distributions are scaled to the mean values of
  the distributions.
}
\label{normalized5}
\end{center}
\end{figure}

Width of the multiplicity distribution for a given centrality gives
the extent of the fluctuation. Thus the physics origin of the fluctuations 
are inherent in the width of the multiplicity distributions. One of
the ways to understand this to plot the multiplicity distributions
within a centrality bin by scaling it to the mean value of
multiplicity ($\langle N_{\rm ch}\rangle$). This is presented in
Fig.~\ref{normalized5} for  Au+Au collisions at \sNN~= 62.4 and 200
GeV using default AMPT and 
Pb+Pb collisions at \sNN~=2.76~TeV using the SM mode of AMPT. 
The vertical axes are multiplied by different factors for better
visibility. In this representation, it is observed that the widths of
the distributions are inversely proportional to volume, that is to 
$\langle N_{\rm ch}\rangle$. Thus the distributions become narrower in
going from peripheral to central collisions for all energies.
This extensive nature of the representation is avoided by calculating
the scaled variance as defined in Eq.~(1).

\section{Multiplicity Fluctuations}

Multiplicity fluctuations are studied as a function of collision
centrality for Au+Au collisions at 
\sNN~=  7.7 GeV,19.6 GeV, 27 GeV,
62.4 GeV, 200 GeV and Pb+Pb collisions for \sNN~=~2.76~TeV for 5\% centrality
bins from peripheral to central collisions. For each centrality bin,
the multiplicity distributions are corrected using centrality bin
width correction method. The AMPT model gives the number of
participating nucleons for each centrality bin and so the results are
presented as a function of \Npart.
The statistical errors of the $\mu$ and $\sigma$ are calculated using the
Delta theorem~\cite{delta} method. Errors for \wch~are obtained by
propagating the errors on $\mu$ and $\sigma$.
In most cases, statistical errors are observed to be small.

\begin{figure}[tbp]
\begin{center}
\includegraphics[height=0.5\textheight,width=0.45\textwidth]{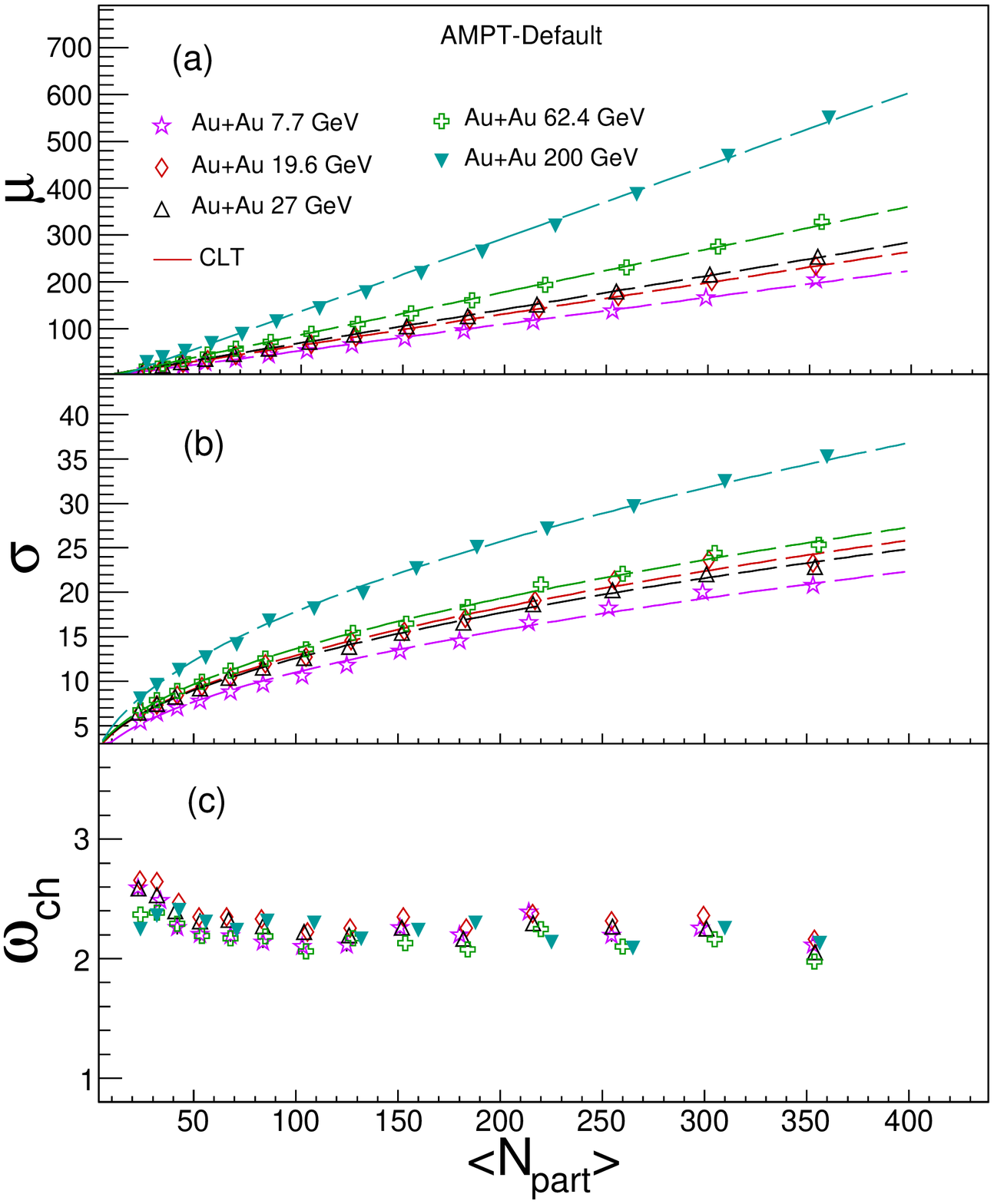}
\includegraphics[height=0.5\textheight,width=0.45\textwidth]{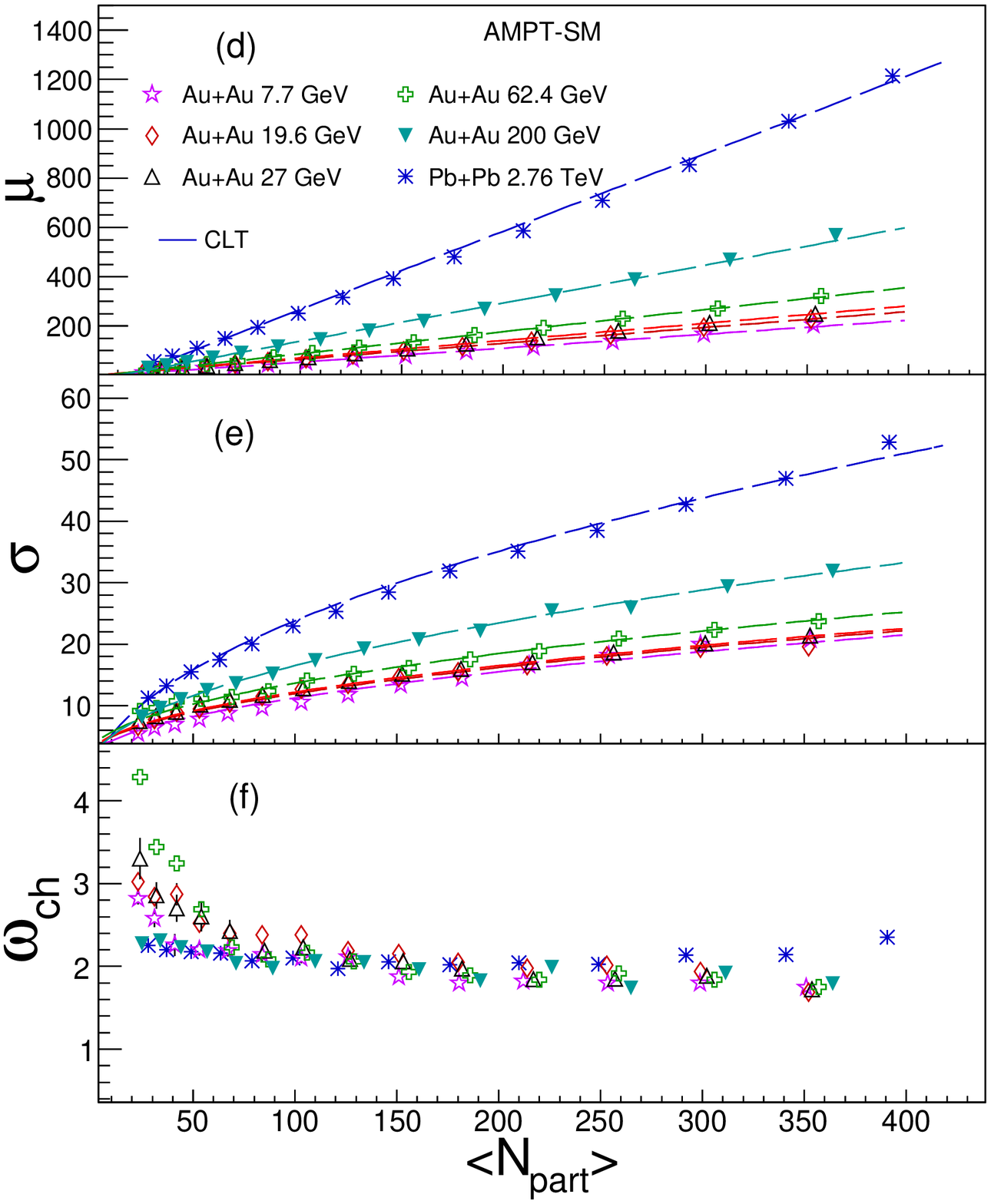}
\caption{(colour online) 
Mean, sigma and 
scaled variance of charged particle multiplicities within  
\acceptance~as a function of centrality for a wide range of 
collision energies. The left panels show the events generated using 
the default mode of AMPT and the right panels show the corresponding 
distributions from the SM mode of AMPT. 
Dashed lines represent fits using the central limit theorem. 
}
\label{fluc1}
\end{center}
\end{figure}

Figure~\ref{fluc1} shows the results for $\mu$, $\sigma$ and
\wch~as a function of $\langle N_{\rm part}\rangle$ for five collision energies.
The left panels show the results for events generated using the
default mode of AMPT and the right panels give the results obtained
using SM mode of AMPT. For \sNN~=~2.76~TeV, only the results from the
SM mode are presented.
It is observed that for all collision energies, $\mu$ and $\sigma$ increase
smoothly in going from peripheral to central collisions for all energies.
The centrality evolution of the 
moments can be understood by the 
Central Limit Theorem (CLT) according to which,
\begin{eqnarray}
\mu\propto\langle N_{\rm part}\rangle \\
\sigma\propto \sqrt{\langle N_{\rm part}\rangle}. 
\end{eqnarray}
It is to be noted that $\langle N_{\rm part}\rangle$ 
is proportional to the volume of the 
system, and so \wch~is a volume independent term. 
In Figure~\ref{fluc1}, $\mu$ and $\sigma$ are fitted with respective 
CLT-form as in the above expressions with the constant of proportionality as free 
parameter.  The CLT curves are superimposed on the AMPT
points. The centrality evolution of the moments follow the trend of
the CLT at all energies. 
Deviations to CLT fits are seen for central collisions 
at the highest energy considered.

The bottom panels of Fig.~\ref{fluc1} show the scaled variances (\wch) 
as a function of centrality for different collision energies. 
The results are similar for both default and SM  
modes of AMPT. At low collision energies, \wch~show a drop in going from most peripheral  
 collisions after which the values remain unchanged. 
At higher energies, \wch~remain rather constant as a function of 
centrality. 

\begin{figure}[tbp]
\begin{center}
\includegraphics[width=0.7\textwidth]{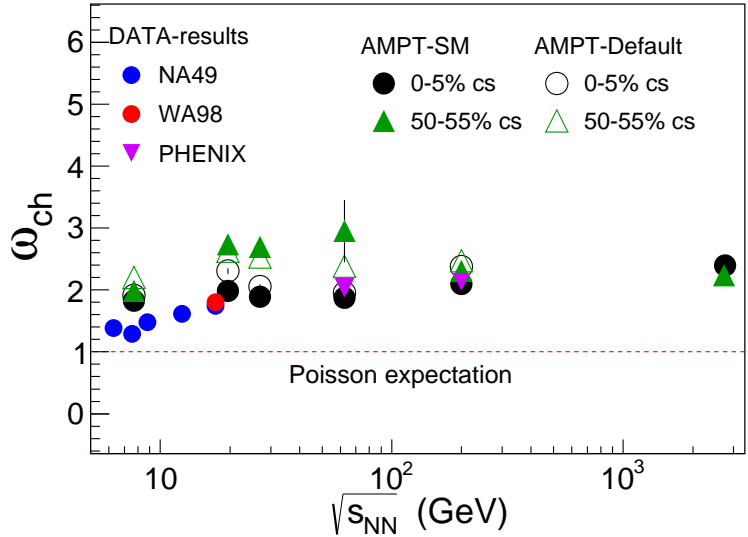}
\caption{(colour online) Beam-energy dependence of scaled variance
  (\wch) of charged particle multiplicity distribution for central
  (0-5\% cross section) and peripheral (50-55\% of cross section)
  collisions as a function of collision energy for
available experimental data and for events generated using two modes of AMPT model.
}
\label{beam_energy_omega}
\end{center}
\end{figure}

Beam-energy dependences of the multiplicity fluctuations have been
studied by combining results from available
experimental data with AMPT model calculations. Experimental results
for heavy-ion collisions are available for WA98~\cite{wa98} and NA49~\cite{na49,na49-Ryb}
 experiments at
CERN SPS and PHENIX~\cite{Phenix} experiment at RHIC.
Since these experimental results are presented for different detector
acceptances, these have to be scaled to a common acceptance in
order to present in the same figure. The available results are scaled
for $\Delta\eta<1$ using the prescription given Ref.~\cite{Phenix}.
If $\omega_{\rm acc1}$ represents the measured scaled variance and
$\omega_{\rm acc2}$ is the scaled variance within $\Delta\eta <1$, then we have~\cite{Phenix},
\begin{equation}
\omega_{\rm acc2}=1+f_{\rm acc}(\omega_{\rm acc1}-1)
\end{equation}
where, 
\begin{equation}
f_{\rm acc}=\frac{\mu_{\rm acc2}}{\mu_{\rm acc1}}.
\end{equation}
The values of $\omega_{\rm acc2}$ have been calculated from the data
provided by the experiments.
Fig.~\ref{beam_energy_omega} shows the values of \wch~for central
collisions for WA98, NA49 and PHENIX experiments.
The results for \wch~are also presented for two 
different centralities (0-5\% and
50-55\% of total cross section) using the default and SM modes of
AMPT. A slow rise in \wch~has been observed from low to high collision
energy and then remaining constant at higher energies.
The AMPT results overestimate those of NA49 experimental data, but
are close to those of WA98 and PHENIX data.
These values are larger compared to the Poisson expectations. 

\section{Discussions}

Collision energy dependence of fluctuations of charged particle
multiplicity, presented in Fig.~\ref{beam_energy_omega} does not show
any non-monotonic behaviour for the AMPT results as well as for
experimental data. The experimental data and AMPT results are rather
close to each other. Non-monotonic behaviour is not expected from the
AMPT event generator as it does not contain any physics specific to
phase transition and critical behaviour. The absence of non-monotonic
behaviour in the experimental data point to the absence of critical
phenomenon for the systems studied at SPS and RHIC. 
In addition,
the observed fluctuations in charged particles may be affected by the
evolution of fluctuation during the early collision time to
freeze-out. More data for Beam Energy Scan (BES) energies at RHIC are
needed to make any definitive conclusion on the critical behaviour. 
The results presented using the AMPT event generator provide baselines for these
studies at BES energies and for collisions at the Large Hadron
Collider (LHC).

Multiplicity fluctuations arise from several known sources such as, 
fluctuations in the number of sources producing multiplicity,
fluctuations in the number of particles produced in each source, 
detector-acceptance, resonance decays, etc. If particles are produced 
independently, one gets $\omega_{ch} = 1$. 
But as we move to higher energies, the 
non-statistical fluctuations increase and automatically 
contribute to the increased value of the fluctuation as discussed in 
Ref.~\cite{danilov}. 
Various studies have been reported in the literature in order to explain the
values for multiplicity fluctuations expressed in terms of the scaled
variances~\cite{marek2,heiselberg,baym,Phenix,begun}. 
Ref.~\cite{begun} gives a prediction for the values of
scaled variance in grand canonical ensemble (GCE), canonical ensemble
(CE) and micro canonical ensemble 
(MCE) using the hadron resonance gas
model at chemical freeze-out for the central heavy-ion collisions for a
wide collision energy range. According to these calculations, we get a
value for $\omega_{ch}$ between 1.4 to 1.64 for GCE, 1.06 to 1.64 for
CE, and 0.534 to 0.619 for MCE. At higher energies, the scaled
variance is predicted to be similar for CE and GCE.
In Ref.~\cite{na49}, it is observed that, the
values for the scaled variance from NA49 experiment is better
described by MCE. Results presented in Fig.~\ref{beam_energy_omega}
are close to the GCE description for higher collision energies.

An estimation for the multiplicity fluctuation can be made in the light
of the participant model, where the nucleus-nucleus collisions are
assumed to be superposition of nucleon-nucleon interactions
(as described in Ref.~\cite{heiselberg}). Here, the total multiplicity
fluctuation has contributions due to fluctuations in $N_{\rm part}$
and also due to the fluctuation in the number of particles produced
per participant. In this formulation, $\omega_{\rm ch}$ can be expressed as, 
\begin{eqnarray}
\omega_{\rm ch} = \omega_{\rm n} + \langle n \rangle \omega_{N_{\rm part}}
\end{eqnarray}
where, $n$ is the number of charged particles produced per participant, 
$\omega_{\rm n}$ denotes fluctuations in $n$, and $\omega_{N_{\rm part}}$ is  the
fluctuation in $N_{\rm part}$. 
The value of $\omega_{\rm n}$ has a
strong dependence on acceptance. The fluctuations in the number of
accepted particles ($n$) out of the total number of produced particles
($m$) can be calculated by assuming that the distribution of $n$ follows a
binomial distribution. This is given as~\cite{heiselberg,wa98},
\begin{eqnarray}
\omega_{\rm n} = 1 - f + f \omega_{\rm m},
\end{eqnarray}
where $f$ is the fraction of particles accepted. The values of $f$ are
obtained from the published proton-proton collision data for total number of
charged particles and number of charged particles in mid-rapidity over the energy range
considered~\cite{phobos,alice1,cms}. 
$\omega_{\rm m}$ is calculated from the total number of
charged particles using the formulation given in
Ref.~\cite{wa98}. Using these, we obtain the values of 
$\omega_{\rm  n}$ as a function of collision energy. 
The values of $\omega_{\rm n}$ vary within 0.98 to 2.0
corresponding to \sNN~=7.7~GeV to 2.76~TeV.
These values match with those reported for SPS
collisions~\cite{wa98,danilov}. 
%By being different from unity at higher energies, the
%quantity, $\omega_{\rm n}$ may contain possible 
%dynamical contribution to $\omega_{\rm ch}$. 
By using the values of $\omega_{\rm n}$ in Eqn.~(10), 
we find that $\omega_{\rm ch}$ from the statistical model calculations
are close to those of the AMPT results presented in Fig.~7. We observe
that the values of $\omega_{\rm n}$ has a major contribution to 
$\omega_{\rm ch}$.

\section{Summary}

We have presented
a comprehensive study on the fluctuations of charged particle
multiplicity at 
mid-rapidity as a function of collision centrality
and beam-energy. We have studied the multiplicity distributions of produced charged particles  
and their event-by-event fluctuations 
for heavy-ion collisions using the available experimental data
and AMPT model calculations.  
We have demonstrated the importance of the choice of centrality 
selection and 
a detailed discussion on the bin-width effect and its remedy have been 
presented. 
The scaled variance,
 $\omega_{ch}$ is constructed in such a way that statistical
 fluctuations give the same result at any multiplicity. Thus, it
has negligible dependence on centrality and beam energy. 
Comparison with experimental data and AMPT results  has been presented after a proper rescaling 
to consider the difference in the geometrical acceptance. 
We observe that the mean and width of the multiplicity distributions 
monotonically increase as a function of increasing centrality at all
collision energies, 
whereas $\omega_{\rm ch}$ shows minimal variation with centrality.
The beam energy dependence shows that $\omega_{\rm ch}$  
has a slow rise at lower collision energies and remain constant at higher energies.
$\omega_{\rm ch}$
at higher energies and central collisions are found to be $\sim$2 within the limits of statistical precision.
The model calculations exhibit good agreement with the results from
WA98, NA49 and PHENIX experiments.
The values of the $\omega_{\rm ch}$ are found to be consistent with the
participant model calculations, where charged particle fluctuation is
considered as a sum of particle number fluctuation from each source
and fluctuation in number of sources. We observe that the 
fluctuation in number of
particles produced per participant has a dominant effect on
$\omega_{\rm ch}$.
Both data and model calculations have no distinct signature of any non-monotonic variation.
Our study offers a baseline for the future endeavour to pursue research on particle multiplicity
fluctuations at  Facility for Antiproton and Ion Research (FAIR), RHIC
and LHC energies.

\bigskip

\noindent
{\bf Acknowledgement}

This research used resources of the LHC Grid computing
facility at Variable Energy Cyclotron Centre, Kolkata.

%\linenumbers
\end{document}